\documentclass[12pt, preprint]{aastex}

\newcommand{\kms}{km s$^{-1}$ }
\newcommand{\schi}{{\sc Hi} }
\newcommand{\schii}{{\sc Hii}\ }
\newcommand{\ms}{M$_{\odot}$ }
\newcommand{\vlsr}{$v_{\rm LSR}$ }
\newcommand{\vexp}{$v_{\rm exp}$ }
\newcommand{\halpha}{H$\alpha$ }
\def\simlt{\lower.5ex\hbox{$\; \buildrel < \over \sim \;$}}
\def\simgt{\lower.5ex\hbox{$\; \buildrel > \over \sim \;$}}

\shorttitle{An Old Supernova Remnant within an HII Complex}
\shortauthors{Kang, Koo, and Salter}

\begin{document}

\title{An Old Supernova Remnant within an HII Complex at $\ell\approx 173^\circ$: FVW 172.8+1.5}
\author{Ji-hyun Kang$^{1,3,4}$, Bon-Chul Koo$^2$, and Chris Salter$^1$}
\affil{$^1$Arecibo Observatory, HC 3 Box 53995, Arecibo, PR 00612}
\affil{$^2$Department of Physics and Astronomy, Seoul National University, Gwanak 599 Gwanak-ro, Gwanak-gu,
Seoul 151-742, Korea}
\affil{$^3$Yonsei University, 50 Yonsei-ro, Seodaemun-gu, Seoul 120-749, Korea}
\affil{$^4$Korea Astronomy and Space Science Institute 776, Daedeokdae-ro, Yuseong-gu, Daejeon 305-348, Korea}

\email{jkang@kasi.re.kr, koo@astro.snu.ac.kr, csalter@naic.edu}

\begin{abstract}
We present the results of \schi 21-cm line observations
to explore the nature of the high-velocity (HV) \schi\ gas at $\ell\sim 173^\circ$.
In low-resolution \schi surveys this HV gas appears as
faint, wing-like, \schi emission 
that extends to velocities beyond those allowed by Galactic
rotation. We designate this feature
FVW (Forbidden Velocity Wing) 172.8+1.5.
Our high-resolution ($3.'4$) Arecibo \schi observations show that
FVW 172.8+1.5 is composed of knots, filaments, and
ring-like structures distributed over an area a few degrees in extent.
These HV \schi emission features are confined within the limits of
the HII complex G173+1.5,
which is composed of
five Sharpless HII regions distributed along a radio continuum loop of size
$4^\circ.4\times 3^\circ.4$, or $\sim 138\ {\rm pc} \times 107$~pc, at
a distance of 1.8 kpc. G173+1.5
is one of the largest star-forming regions in the outer Galaxy.
We demonstrate that the HV \schi gas is well correlated with
the radio continuum loop and that the two seem to trace an
expanding shell. The expansion velocity of the shell is large (55 \kms)
suggesting that it represents a supernova-remnant (SNR).
We derive physical parameters for the
shell and show these to be consistent with the object being a SNR.
We also detect hot X-ray emitting gas inside the HII complex
by analyzing the ROSAT all-sky X-ray background survey data.
This also supports the SNR interpretation.
We conclude that
the HV \schi gas and the X-rays are most likely the products of
a supernova explosion(s) within the HII complex, possibly in a cluster
that triggered the formation of these HII regions.

\end{abstract}

\keywords{supernova remnants --- ISM: individual (\object{FVW172.8+1.5}) ---
radio lines: ISM --- ISM: {\sc Hii} regions -- stars: formation}


\section{Introduction}

In large-scale ($\ell$,$v$) diagrams of {\sc Hi} 21-cm line emission in
the Galactic plane, there are many small, faint, high-velocity
``wing-like" features extending to velocities well beyond the maximum
or minimum permitted by Galactic rotation \citep{kk04}.
These ``Forbidden Velocity Wings (FVWs)" are likely due to energetic
phenomena in the Galaxy, as they are confined to small areas ($\lesssim
2^\circ$), and project smoothly beyond the general Galactic emission to
high velocities.  \citet{kk04} suggested that some of these FVWs may represent
the expanding shells of ``missing" supernova remnants (SNRs), which are
not included in existing SNR catalogs.
The basic idea is that old SNRs are faint in the radio continuum,
making it difficult to identify them because of both the
confusion due to the Galactic background emission, and 
observational limitations (e.g., Brogan et al. 2006). \citet{kk04}
considered the fact that an old SNR should possess a fast-expanding
\schi shell that will still be present as a coherent entity after the remnant becomes too
faint to be visible in the radio continuum.
If its expansion velocity
is greater than the minimum or maximum velocities permitted by
Galactic rotation, then an old SNR shell, or part of it (maybe the
caps), could be detected as high-velocity \schi gas, e.g., FVWs.
Indeed, \citet{kks06} have carried
out high-resolution \schi line observations toward
one FVW and detected a rapidly expanding
(80 \kms) \schi shell, the parameters of which are
consistent with those of the remnant of a SN that exploded
some 0.3~Myr ago.

Recently, \citet{kk07} identified 87 FVWs from the Leiden/Dwingeloo \schi survey
\citep[LDS;][]{hb97} and the Southern
Galactic Plane Survey data \citep[SGPS;][]{sgps}. Among these, 6 FVWs
are found to be coincident with SNRs, 4 with nearby galaxies, and 3 with
high-velocity clouds.
The rest (85\%) are not associated with any obvious
objects that could be responsible for their high velocities.
We have since been making follow-up high-resolution \schi observations
of these FVWs of unknown nature
using the Arecibo 305-m and Green Bank 100-m radio telescopes
in order to identify their natures and origins, and
have now observed $\sim30$\% of them.
We find that about 40\%
of the observed FVWs have shell-like morphologies. More than half of these 
are apparently expanding at $\simgt 50$~\kms, which supports the possibility of
some FVWs being old SNRs.
The other 60\% show irregular structures consisting
of filaments and clumps. Most of these show
faint bumps in their line profiles indicating high-velocity \schi clouds.
They might correspond to the fast-moving, compact clouds in
the disk, or in the disk-halo interface,
that have been recently detected in sensitive, high-resolution
\schi surveys \citep[][and references therein]{stanimirovic06, begum10}.
The full results from our survey will be presented in a
separate paper.

Here, we report Arecibo \schi 21-cm line observations of
FVW172.8+1.5. This particular FVW is unique in
that it is associated with
an HII complex in the outer Galaxy. We show that
FVW 172.8+1.5 is most likely an old SNR produced in this complex.
The catalog of \citet{kk07} lists this
feature as two objects, FVW173.0+0.0 and FVW173.0+3.0.  However, the
Arecibo \schi image shows that these are likely to be parts of a single
coherent object, which we will call FVW172.8+1.5.  In Section 2, we
describe the \schi observations.  The \schi results
are presented in Section 3. In Section 4, we present
a multi-wavelength view of the HII complex
in this region and investigate its association with FVW 172.8+1.5.
We discuss the origin of FVW172.8+1.5 and
the star-formation history in this area in Section 5,
presenting a summary in Section 6.

\section{The \schi 21-cm Line Observations}

\schi 21-cm line observations of FVW172.8+1.5 were made in 2006
October with the Arecibo 305-m telescope using the 7-beam,
dual-polarization, Arecibo L-band Feed Array (ALFA).  The beam-to-beam
spacing for the scanning pattern used was $1~\!\arcmin\!.8$.  GALSPECT,
a dedicated spectrometer for Galactic \schi ALFA surveys, was used as
the backend.  Both wide- and narrow-band spectra were acquired
\citep[see][for details]{stanimirovic06}.  Here we present data from the narrower
7.14-MHz band, which is analyzed into 8192 equally spaced channels,
giving an unsmoothed velocity resolution of 0.18~\kms$\!$.

The field observed covers an area of $7~\!\degr\!\!.5 \times
4~\!\degr\!\!.5$ centered at $(\alpha, \delta) = (5^{\rm h} 37^{\rm m},
35~\!\!\degr 40~\!\arcmin$) (J2000).  The observations employed the
basket-weaving technique \citep{peektech, peek07}, which scans the sky with a
zig-zag pattern by driving the telescope up and down on the prime
meridian at a speed of 1.5 arcmin/sec.  For bandpass correction, a
``Least-Squares Frequency Switching'' \citep[LSFS;][]{lsfs} calibration
was performed each day.  Data reduction used the IDL pipe-line codes
developed by the Berkeley group \citep{peektech}.  Using these, the
data are converted into a brightness temperature cube. 
The contamination due to the coma sidelobes of the ALFA off-centre beams
has been corrected \citep[see][for the details of data reduction]{peektech}. 
The final cube
has a velocity resolution of 0.74~\kms, a spatial HPBW of 3.4 arcmin,
and an rms of 0.13 K.

\section{The \schi Results}

\begin{figure}
\epsscale{1.0}
\plotone{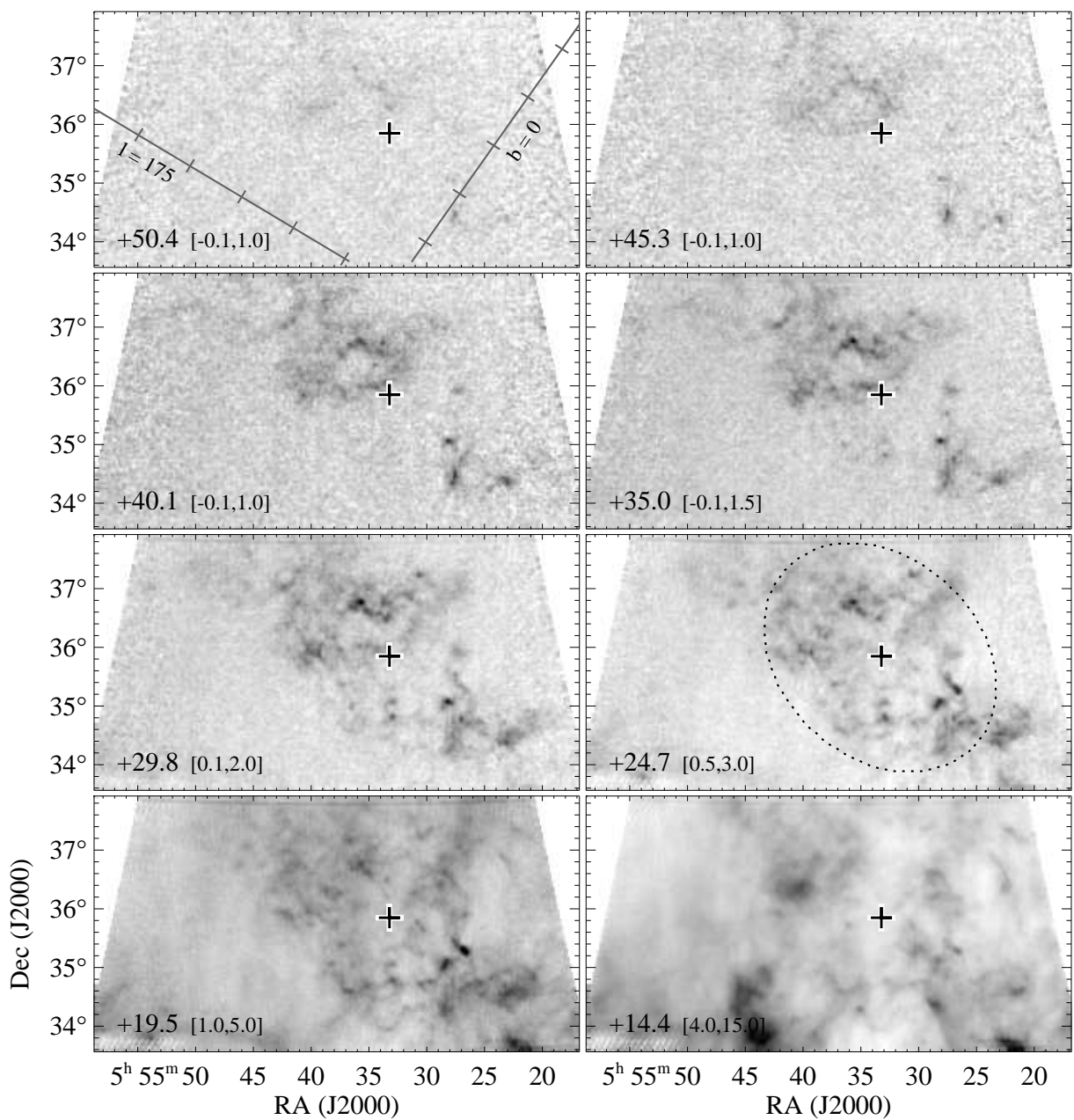}
\caption{\schi channel images of the observed field.
Each image is integrated over a velocity interval of 5.1 km s$^{-1}$.
The central LSR velocity
is given at the bottom left of each image.
The numbers in the brackets are the minimum and maximum
brightness temperatures (K) of the grey scale which is linear.
The black cross indicates the approximate center of
the {\sc Hi} structure. The dotted ellipse on
the +24.7 \kms\ frame marks the boundary of the proposed shell
as adopted in this paper.
\label{fg1}}
\end{figure}

Fig.~\ref{fg1} shows channel images of the Arecibo \schi data.
The \schi emission associated with FVW172.8+1.5 is visible
for +19.5~\kms $\la$ \vlsr $\la$ +50.4~\kms.
At the highest velocities (+40.1~\kms $\la$ \vlsr $\la$ +50.4~\kms),
the {\sc Hi} appears to be largely separated
into two concentrations 
centered at $\sim (5^{\rm h} 37^{\rm m}, +36\degr
30\arcmin)$ and $\sim (5^{\rm h} 27^{\rm m}, +34\degr 30\arcmin)$,
These were previously designated FVW173.0+3.0 and FVW173.0+0.0 respectively.
The northern concentration (FVW173.0+3.0) looks diffuse and clumpy,
and appears to form a ring-like structure with a diameter of
$\sim 2\!$~\degr.
In contrast, the southern concentration (FVW173.0+0.0)
appears to be confined to a few small ($\la 30^\prime$) regions.
At lower velocities (+16.6~\kms $\la$ \vlsr $\la$ +35.0~\kms),
a thin, filamentary feature that surrounds the two concentrations
appears, so that the entire structure appears to be a large shell
with an extent of $4\degr.4\times3.4\degr$.
The crosses in Fig.~\ref{fg1}
and the ellipse superposed on the +24.7 \kms\ channel map mark the
approximate center and boundary of the shell structure.
At +14.4 \kms\ and lower velocities, the \schi emission from the Galactic ISM dominates,
so that no particular \schi feature associated with the source
is recognizable (see also \S~4.1.1).

\begin{figure}
\epsscale{1.0}
\plotone{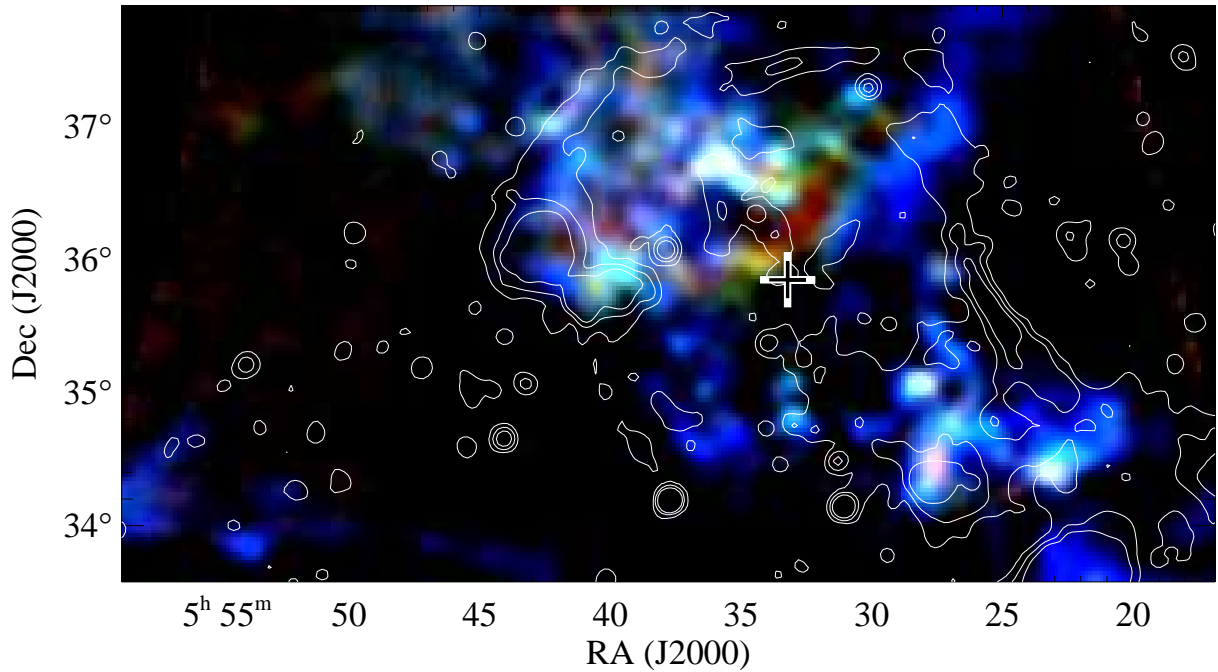}
\caption{A three-color image showing the Arecibo \schi  emission from
FVW172.8+1.5. Red, green, and blue represent the images integrated over LSR
velocities of +45 to +35, +35 to +25, and +25 to +20 km s$^{-1}$ respectively.
Effelsberg 11-cm radio continuum contours \citep{furst} are overplotted.
Contour levels are 30, 100, 200 mK in brightness temperature.
\label{fg2}}
\end{figure}

The various \schi features described above are also visible
in Fig.~\ref{fg2}, which is a  three-color image of
the \schi gas integrated over different velocity intervals.
In this figure, we superpose
the Effelsberg 11-cm radio continuum contours \citep{furst}. It is seen
that the high-velocity \schi emission features
lie essentially within the radio continuum filaments.
The northwestern high velocity \schi gas,
represented by red and green colors, are
confined within the~ northern part of
the continuum structure, while
the \schi features at lower velocities (colored blue)
lie along the inner boundary of the outer radio continuum filaments.
The morphological relation between the \schi and the radio filaments
strongly suggests their association. In addition, there is little possibility of 
chance alignment of the two structures, because confusion is low
along the line of sight in the direction of $\ell\sim 173^\circ$.
The increase of the surface area and/or the size of the \schi-emitting
region with decreasing velocity suggests that we are looking at 
the receding portion of an expanding shell. 
The \schi morphology does not exactly match
that of a uniformly expanding spherical shell, but this could be
due to a non-uniform and inhomogeneous ambient interstellar medium.
The radio continuum structure is part of the HII complex
in this area, and we will investigate the association of the two
at other wavelengths in Section 4 and
their physical properties and origin in Section 5.

\begin{figure}
\epsscale{1.0}
\plotone{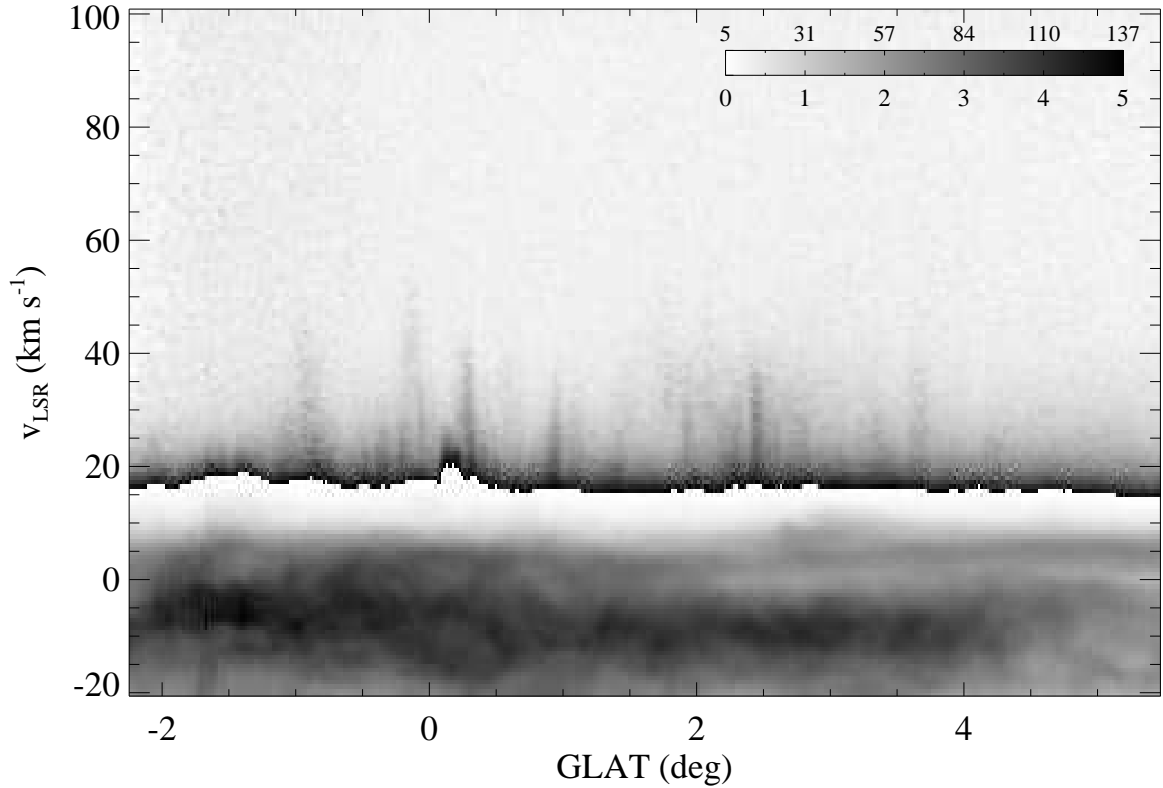}
\caption{
The \schi peak brightness temperature over the range of Galactic longitude,
171.\arcdeg8 to 173.\arcdeg7, as a function of Galactic latitude.
In order
to show \schi features at lower velocities, the intensity scale for
brightness temperatures in excess of 5 K is greatly expanded.
The grey-scale ranges are shown by the bar in the upper right of the
figure; values for brightness temperatures $<5$ K are marked below the
bar, with those for $>5$ K are marked above. The merit of a peak
intensity image is that it enhances bright structures that would not be
seen in an averaged image.
\label{fg3}}
\end{figure}

\begin{figure}
\epsscale{1.0}
\plotone{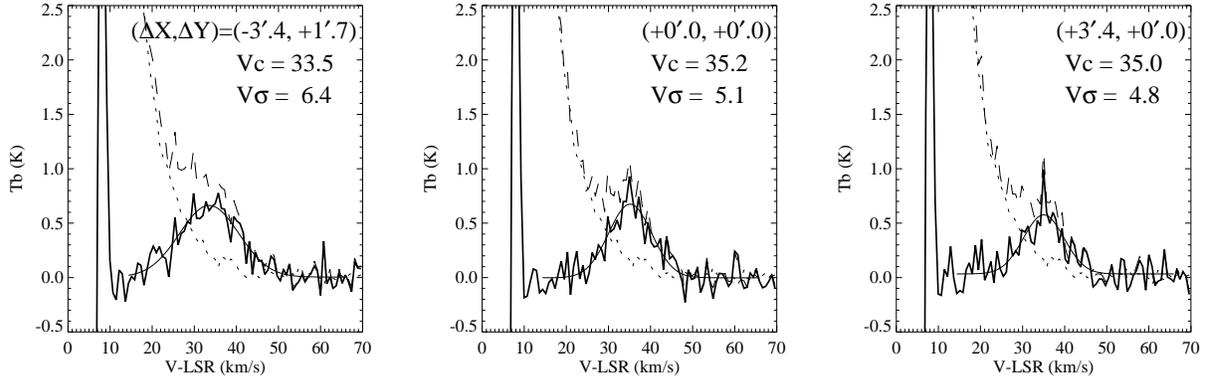}
\caption{Sample line profiles in the area around
(5$^{\rm h}$ 34.5$^{\rm m}$, +36$\degr$ 03$'$).
The relative positions are shown in each frame.
Dashed and thick solid lines show the original and background-subtracted
line profiles,
respectively. The thin solid lines show the best fit of Gaussian profiles
to the background-subtracted profiles. The central velocities and velocity dispersions
of the Gaussians in \kms\ are given in each frame.
\label{fg4}}
\end{figure}

The velocity structure of the high-velocity \schi features is
shown in Fig.~\ref{fg3}, which shows
the peak \schi brightness temperature in Galactic longitude
as a function of Galactic latitude and velocity.
This position-velocity diagram shows the high-velocity gas over the
entire field in a single diagram and enhances
the appearance of small-scale wing-like features.
Fig.~\ref{fg3} shows that FVW172.8+1.5 is composed of many
small-scale, high-velocity wings spatially confined to
$-1\degr < b < +4\degr$, which suggests an inhomogeneus nature
for the shell. These wings are mostly
straight but some have the shape of an extended ring,
e.g., that at $b=-0.2\degr$. This indicates that
the shell is composed of small clumps.
Since only
the high-velocity end portions of the wings are visible, their
central velocities are usually not accessible.
However, a few line profiles in the area near
$(5^{\rm h} 34.5^{\rm m}, +36\degr 03\arcmin)$
show discrete spectral features at high velocities (Fig.~\ref{fg4}).
A Gaussian fit to these features
results in a central velocity of +33~--~+35~\kms and a
velocity dispersion of 4.3 -- 6.4~\kms, corresponding to 
kinetic temperatures of 2,200 -- 4,900 K if the broadening is 
entirely due to thermal motions.

\section{The $\ell=173\degr$ region in Multi Wavebands}

\subsection{Radio Continuum}
\subsubsection{The HII Complex G173+1.5}

\begin{figure}
\epsscale{1.0}
\plotone{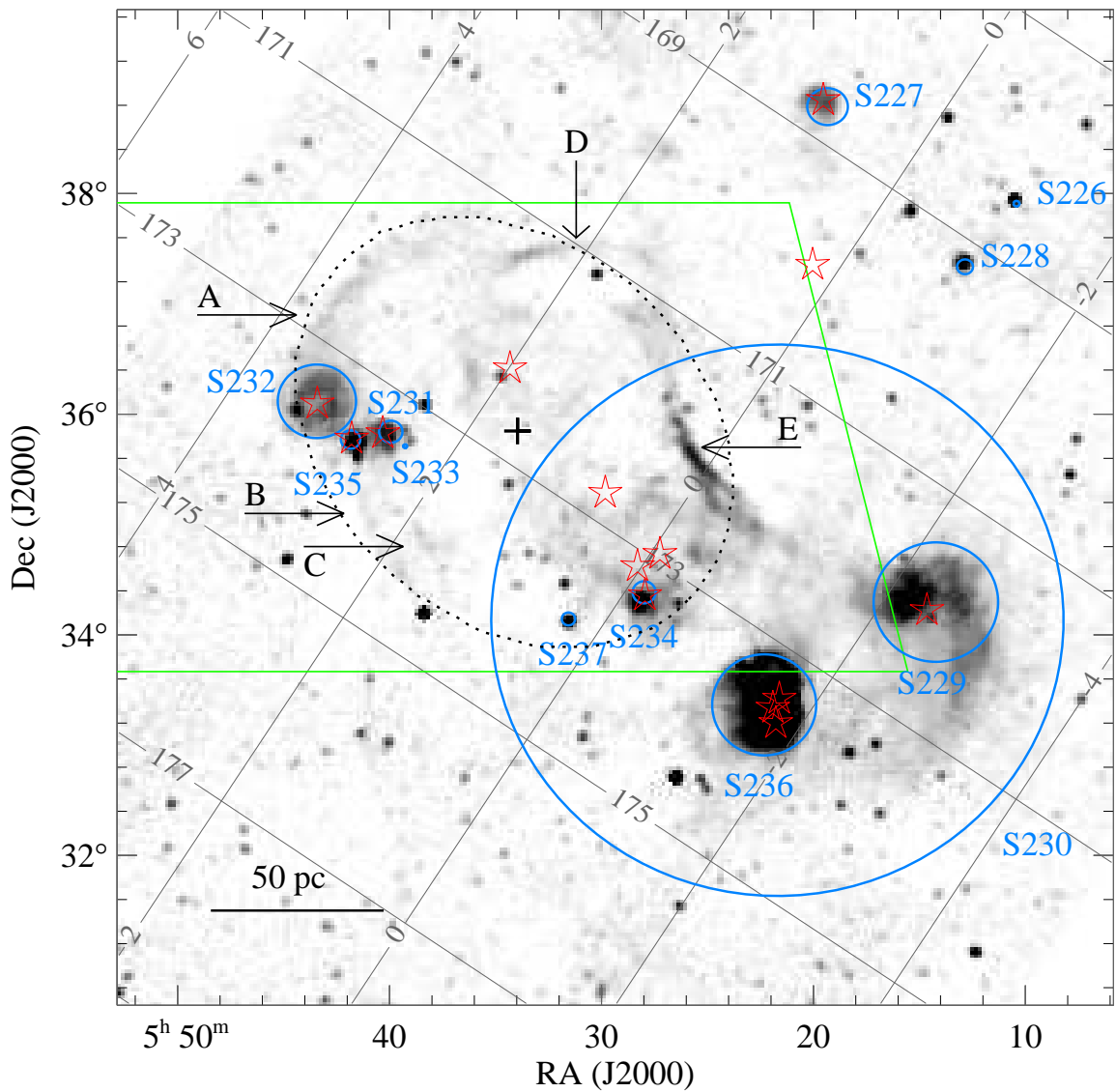}
\caption{
A grey-scale plot of the Effelsberg 11-cm radio continuum in the direction of
FVW172.8+1.5.
Sharpless \schii regions are marked by blue circles, while
 Galactic O stars are marked by red star symbols.
The region observed at Arecibo (Fig. 1) is shown by green lines.
The black cross and the dotted ellipse are the approximate center
and boundary adopted for the shell, respectively. The continuum filaments
defining the \schii complex are marked by arrows. The scale bar in the 
bottom left represents 50 pc at 1.8 kpc.
\label{fg5}}
\end{figure}

The region of Auriga within which FVW172.8+1.5 lies is an HII complex
composed of Sharpless HII regions and OB associations
in the Perseus arm (Fig.~\ref{fg5}).
The HII regions are organized along a large
($\sim 7^\circ \times 4^\circ$)
filamentary structure resembling a bow tie \citep{kerton07}.
The \schii regions have been studied previously
\citep[e.g.][]{israel78}, and they are largely at two different
distances (see Table~\ref{tbl-1}).
To the northeast of the structure, S231, S232, S233, and S235
are known to be associated with a giant molecular cloud
at a distance of 1.8~kpc \citep{evans81},
with CO radial velocities
of $-18.1$ to $-23.0$~\kms$\!$ \citep{blitz82},
in which active star formation is on-going
(see \S~4.3).
The distances determined by
spectrophotometry of the exciting stars of individual HII regions
range from 1.0 to 2.3 kpc, but the distance to the associated
molecular cloud (1.8 kpc) is often adopted as the distance to
the whole HII complex.
To the south of the complex, two $1\degr$-sized HII regions
S229 (IC 405) and S236 (IC 410) and the large ($\sim 5\degr$)
diffuse HII region S230 are present.
S236 is associated with a molecular cloud at $-7.2$~\kms\ and
might be at a considerably greater distance
because the estimated distance to its central star cluster (NGC 1893) 
ranges from 3.2 to 6 kpc
\citep[][and references therein]{sharma07}.
For the other two HII regions (S229 and S230), not much has been known, although
\citet{fich90} measured H$\alpha$ velocities of
+4.4 and  0.0 \kms\ toward S229 and S230, respectively.
Toward S229, CO $J=1$--0 emission at $+6.7$ \kms has been detected, but
it may not be related to the HII region \citep{blitz82}.
The small HII regions
S234 (IC 417) and S237 in the middle of the ``bow-tie''
have velocities
of $-13$ and $-4$~\kms, similar to the first and second groups
respectively.
A recent photometric determination of the
distance to the central cluster (Stock 8) of S234
also agrees with the distance to the northwestern group
of HII regions, i.e., $2.05\pm 0.10$ kpc \citep{jose08}.
We therefore consider that the Sharpless HII regions S231--S235
form the active star formation complex G173+1.5 at
a distance of 1.8 kpc at \vlsr$\approx -20$~\kms.
The other HII regions have considerably different velocities and
distances, and they are not considered to be associated with this complex.

\begin{figure}
\epsscale{0.6}
\plotone{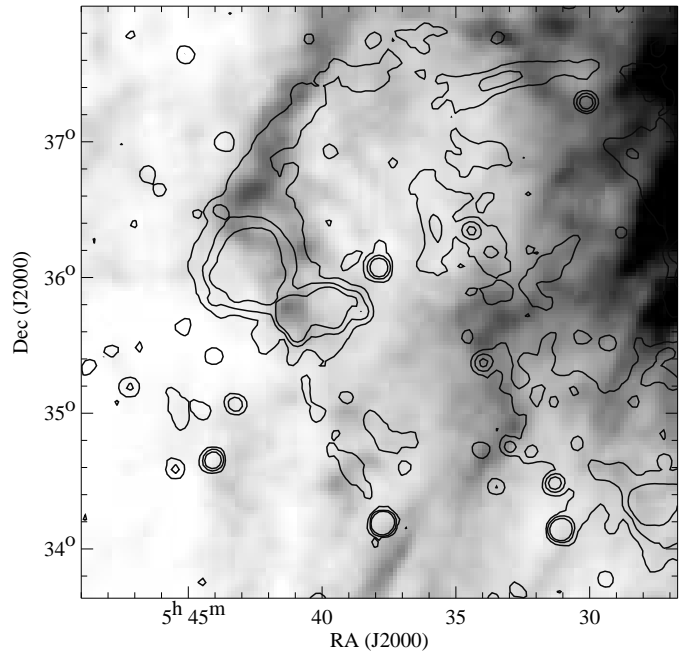}
\caption{An \schi map
integrated from \vlsr$=-28$ to $-25$ \kms, with the 11-cm continuum contours overlaid.
An HI filament along the continuum filament A is visible (cf. Fig. 5),
although the correlation is not perfect.
\label{fg6}}
\end{figure}

The morphology of the region suggests an association between the large radio
continuum filamentary structures
and the HII regions.
Filaments A through D (see Fig.~5) have similar
concave shapes suggesting that they have resulted from a
common source within the HII complex.
For the radio filament A, we
have found an HI filamentary
structure along the radio feature
at \vlsr=$-25$ to $-28$ \kms (Fig.~\ref{fg6}).
The correlation is not perfect in the sense that
the HI feature looks less curved than the continuum filament
and seems to extend outside of the field shown in Fig.~6.
However, the velocity is close to the CO velocity of S232,
and the two could be associated.
We could not find any 
\schi structures associated with the other continuum filaments.
Filament E has an enhanced brightness and
its curvature is opposite to those of Filaments A--D.
It is likely that the enhanced brightness
and the convex shape is due to interaction with
a dense ambient medium there, although no responsible
molecular cloud is seen
in the available CO survey data \citep[][see also Fig. 9]{dame01}.
Hence, filaments A--E are likely to be associated with
the HII complex G173+1.5, although we have only circumstantial
evidence, except perhaps for filament A.  Kinematic evidence
is needed to confirm the association.

\subsubsection{The Radio Continuum Spectrum of Filament A}

The nature of the radio emission from the
large continuum filamentary structure
has been discussed previously, although not in detail.
\citet{kerton07} noted this structure
in the 1420 MHz Canadian Galactic Plane Survey map, and argued that
the filaments D \& E are thermal emission based on either
the presence of corresponding infrared emission and/or the rising
spectrum between 408 and 1420 MHz.
\citet{gao10} claim that filament A shows a thermal
spectrum between the higher frequencies of 1.4 and 5 GHz.

\begin{figure}
\epsscale{0.8}
\plotone{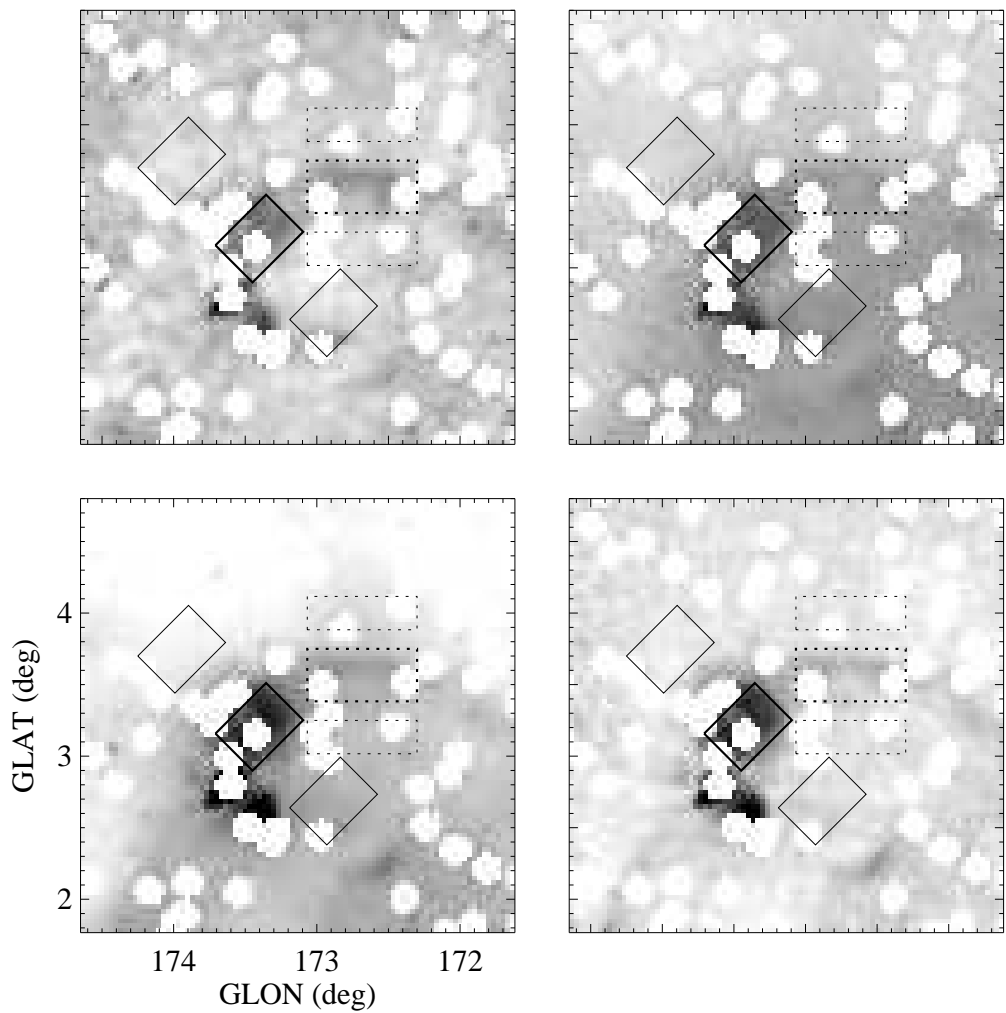}
\caption{
WENSS 325 MHz (top left), CGPS 408 MHz (top right),
CGPS 1420 MHz (bottom left), and Effelsberg 2695 MHz (bottom right)
radio continuum images of the northern part of the FVW173+1.5 field. The areas used
to derive the total flux density and to plot T-T diagrams of radio continuum filament
A and the \schii region, S232, are marked by thick dotted and solid boxes,
respectively. The neighboring areas used for the subtraction of the large scale
background are marked by thin dotted and solid boxes.
The white holes in the images are where point sources have been subtracted.
\label{fg7}}
\end{figure}

Using radio continuum data from 325 MHz to 2.7 GHz
available on the web (Table.~\ref{tbcont}),
we have attempted a study of the continuum spectrum
of the filaments associated with the structure.
The faint filamentary features are visible at most of these frequencies.
Here we will focus on Filament A at $(\ell, b) \sim (172^{\circ}\!\!.5,
+3^{\circ}\!\!.5)$, which stands out at 325 MHz in WENSS.
To estimate values of spectral index,
we (1)
converted the WENSS data to brightness temperature, thus bringing all surveys
to the same intensity units,
(2) convolved all images to have a circular beam of FWHM~=~4\arcmin.7,
the declination beam size of the CGPS 408 MHz image at $\delta=37$\degr,
(3) gridded the images onto the $2'$-spaced pixels of the Effelsberg
data, and 
(4) subtracted out all point sources.
The final images are shown in Fig.~\ref{fg7}, where
the dotted and solid boxes mark
Filament~A and the \schii region, S232,
which provides a reference with a thermal spectrum.
Finally, we removed a smooth, linear,
large-scale background emission gradient using two
neighboring areas, marked by pairs of boxes flanking
the \schii region and the filament in Fig.~\ref{fg7}.

\begin{figure}
\epsscale{1.0}
\plotone{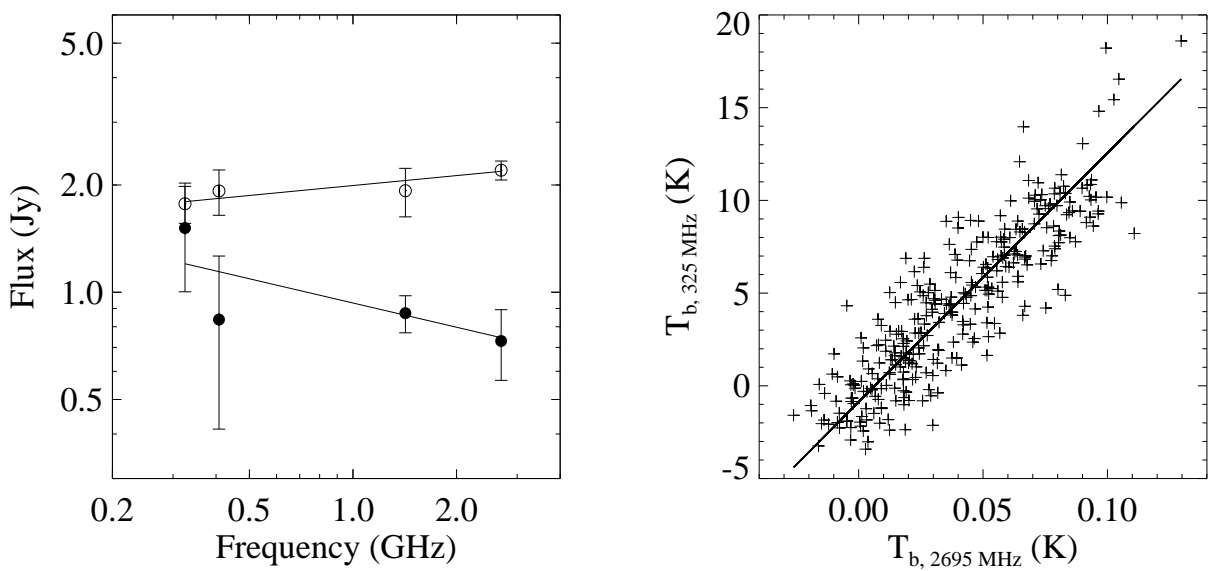}
\caption{(Left) Flux densities of the \schii region S232 (open circles)
and the continuum filament A (filled circles).
The fitted spectra, with $\alpha$ ($S \propto \nu^{-\alpha}$) $= -0.09 \pm 0.02$
and $+0.23 \pm 0.06$ for the \schii region and filament respectively, are
overplotted.
(Right) The T-T plot for the area of filament A using
the WENSS 325~MHz and Effelsberg 2695
MHz data. The best linear fit with $\alpha = +0.32 \pm 0.06$ is overplotted.
\label{fg8}}
\end{figure}

The continuum spectra of Filament A and S232 are shown in
Fig.~\ref{fg8} (left) by filled and open circles respectively. The
total flux densities are given in Table~\ref{tbtotflux}. The error bars
in Fig.~\ref{fg8} represent
standard deviations in the neighboring areas after removal of smooth backgrounds.
Using all flux density values from 325 to 2695 MHz,
the estimated spectral index ($S_\nu \propto \nu^{-\alpha}$)
of S232 is $\alpha = -0.09\pm0.02$, while that of Filament~A is
$\alpha =+0.23\pm0.06$.
We also derived spectral indices for Filament A and S232 using
T-T plots \citep{turtle}. 
The regions  used for making the T-T plots are those
over which the total flux densities were derived.
Fig.~\ref{fg8} (right) shows a T-T plot
between 325 and 2695~MHz of Filament A.
The points are well described by a linear fit.
The slopes of
the plots, and the values of $\alpha$ derived between different pairs of
frequencies, are summarized in Table~\ref{tbtt}.
For S232, the derived spectral indices are consistent, ranging from
$\alpha=-0.04$ to $+0.09$ which, together with the total flux
density spectral
index of $\alpha=-0.09$, is reasonably close
to the value of $\alpha = +0.1$ expected for
an optically thin thermal source.
However, the spectral indices derived for Filament A are not straightforward to
interpret. The frequency pairs that include 325 MHz result in
$\alpha = +0.50\pm0.11$ and $+0.32\pm0.06$, consistent with
non-thermal emission, while the pairs that include 408 MHz
yield $\alpha = +0.07\pm0.10$ and $-0.04\pm0.08$,
compatible with thermal emission.
These trends are consistent with the plot of the total flux densities
(Fig.~\ref{fg8} left).
In that plot, if either the 325 MHz or the 408 MHz data
point is excluded, then the spectral index of Filament~A
would respectively be flatter or steeper than the $\alpha=+0.23\pm0.06$
derived using data at all frequencies.

As mentioned at the beginning of this section,
previous studies claimed that the radio emission
from some of these filaments is of
thermal origin. However, our analysis in this section
indicates the presence of a non-thermal component in filament A.
The fact that the spectral index of S232, as derived from the same
data, is consistent with optically-thin thermal emission
supports our result.
Therefore, it is more likely that the filament has both thermal
and non-thermal components (see also \S~4.3).
If thermal and non-thermal components co-exist in the filament,
a steep spectrum at lower frequencies,
and a flat spectrum at higher frequencies (as found by Gao et al.),
might be expected. Also, if thermal and non-thermal components
co-exist in the continuum emission of Filament~A,
this could be the case for Filaments~B--E. 
We have examined the polarization maps of this area.
The G173 complex region was covered by the polarization surveys 
at 1.4 and 5 GHz \citep{landecker10, gao10}. While the 1.4 GHz 
polarization data of Landecker et al. do not show any apparent 
counterparts to the G173 structure, the 5 GHz polarization data of 
Gao et al. do show some filamentary features associated with the 
structure. In the polarized intensity map of Gao et al. 
(their Fig. 17), two depolarized filaments, indicating a thermal 
nature, are recognizable; one at $(\ell,b) \sim (174.3, +2.0)$, which 
corresponds to Filament B, and the other at $\sim (171.0, +2.5)$, 
located between Filaments A and D. Other than B, the continuum filaments  
have no apparent counterparts in this image. Hence,  
it is difficult to discuss the nature of Filaments A - E using the 
current polarization data, with the exception that a thermal component 
appears to dominate in 
Filament B. The nature of the continuum filaments needs to be 
investigated with more sensitive full-Stokes continuum data.

\subsection{X-rays}

\begin{figure}
\epsscale{1}
\plotone{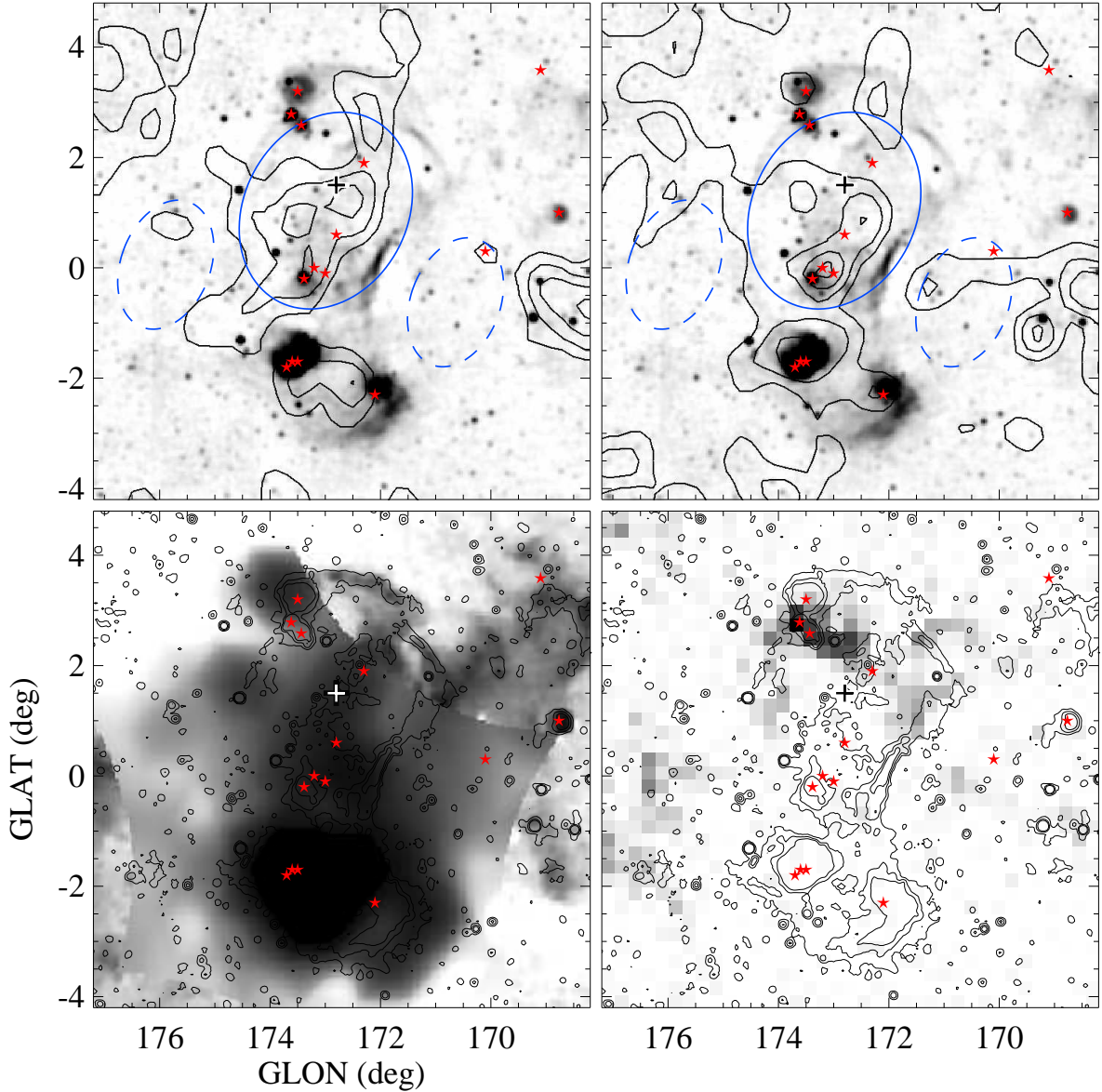}
\caption{(Top left) The ROSAT R45 (3/4 keV) band diffuse X-ray
contour map overplotted on the Effelsberg 11-cm image. The contour levels
are 70, 85, and 100 in counts s$^{-1}$ arcmin$^{-2}$. (Top right)  As for the
top left frame,
but for the ROSAT R67 (1.5 keV) band. The contour levels are 100, 120, 140,
and 160 in counts s$^{-1}$ arcmin$^{-2}$.
(Bottom left) The H$\alpha$ map ($6'$ FWHM resolution)
composite from the Virginia Tech Spectral line Survey (VTSS) and
the Wisconsin H-Alpha Mapper survey \citep{finkbeiner03}. The scale for
the image is 40 -- 70 Rayleigh, but a histogram equalization method is performed
to enhance the contrast of the H-alpha filaments.
(Bottom right) CO J=1-0 intensity image integrated
over \vlsr = $-25$ to $-15$~\kms.
The scale for the image is 0 -- 0.3 K, and is linear.
The blue ellipses in solid and dotted lines overplotted on the top two images
are the area used to measure the X-ray counts
of the shell and the backgrounds, respectively.
The O stars are marked by red stars.
\label{fg9}}
\end{figure}

In the {\em ROSAT Survey Diffuse X-ray Background Map} \citep{snowden97},
there is faint, extended X-ray emission
projected against the HII complex G173+1.5.
(Fig.~\ref{fg9}, top frames).
The emission is mainly from two regions, one within the
S231-235 complex and the other associated with S229/S236.
The emission is not visible in the soft band
(1/4 keV) but appears at the hard bands (3/4 and 1.5 keV).
This is consistent with the emission being associated
with G173+1.5 because if we adopt
$A_V=1.2$ mag (or $N(H)=2.3\times 10^{21}$ cm$^{-2}$)
corresponding to the extinction to S234 \citep{jose08}
as the extinction to G173+1.5,
the transmission of 1/4 keV photons is essentially zero
($\simlt 10$\%)
while it is 40\% and 80\% at 3/4 and 1.5 keV, respectively
\citep{seward00}.
The emission is not uniform and the ratio of 1.5 to 3/4 keV
intensities vary over the field.
Note that,
since the contributions from point sources have been removed from the map,
the emission is from diffuse sources, although some compact sources
might have been included, e.g., the bright spot near S234.
For the present analysis,
we simply derive the X-ray photon counts inside the solid ellipse
marked in Fig.\ref{fg9}. The background is estimated from the surrounding regions
marked by the two dotted ellipses.
The derived count rates over 8.8 deg$^{2}$ are $0.76\pm 0.08$ cts s$^{-1}$ and
$0.78\pm0.11$ cts s$^{-1}$ in 3/4 keV and 1.5 keV bands,
respectively.
The ratio of 1.5 keV to 3/4 keV is $1.0\pm0.2$.
Using the model of
\citet{snowden97}, the ratio corresponds to either
a power-law photon spectrum ($E^{-\alpha}$, where $E$ is energy)
of index $\alpha\sim 4$ or
thermal spectrum with temperature of $\sim 10^{6.9}$~K
for $N(H)=2.3\times 10^{21}$ cm$^{-2}$.
The power-law index $4$ is considerably steeper than that
of a pulsar wind nebula, e.g., $\sim 2$ for the Crab nebula,
so that we may rule out a non-thermal origin.
The diffuse extended morphology also supports a thermal origin.
According to \cite{snowden97}, the attenuated count rate
of thermal X-ray emission from $10^{6.9}$~K gas at 3/4 keV is
$\sim 0.01$ cts s$^{-1}$ acrmin$^{-2}$ for an emission measure of 1 cm$^{-6}$ pc.
Therefore, the observed average 3/4 keV-band
count rate of $2.4\times 10^{-5}$ cts s$^{-1}$ arcmin$^{-2}$
implies an average emission measure of $0.0024$ cm$^{-6}$ pc.
If we use $4R/3$, where $R=60$ pc is the geometrical mean radius of the shell,
as the mean depth along the line of sight,
the mean electron density is about $5.5\times 10^{-3}$ cm$^{-3}$.
This implies a thermal energy for the X-ray emitting gas
of $\sim 3\times 10^{50}$~ergs, assuming that it fills the entire interior of the shell.
If instead the X-ray emitting gas
fills only parts of the shell interior, then the thermal energy will be
somewhat smaller than this.
The derived thermal energy is close to that of an old
SNR, which, together with the high temperature, suggests that the hot
gas might have originated from an SN explosion within the HII 
complex G173+1.5.

\subsection{H$\alpha$ and CO}

\begin{figure}
\epsscale{1}
\plotone{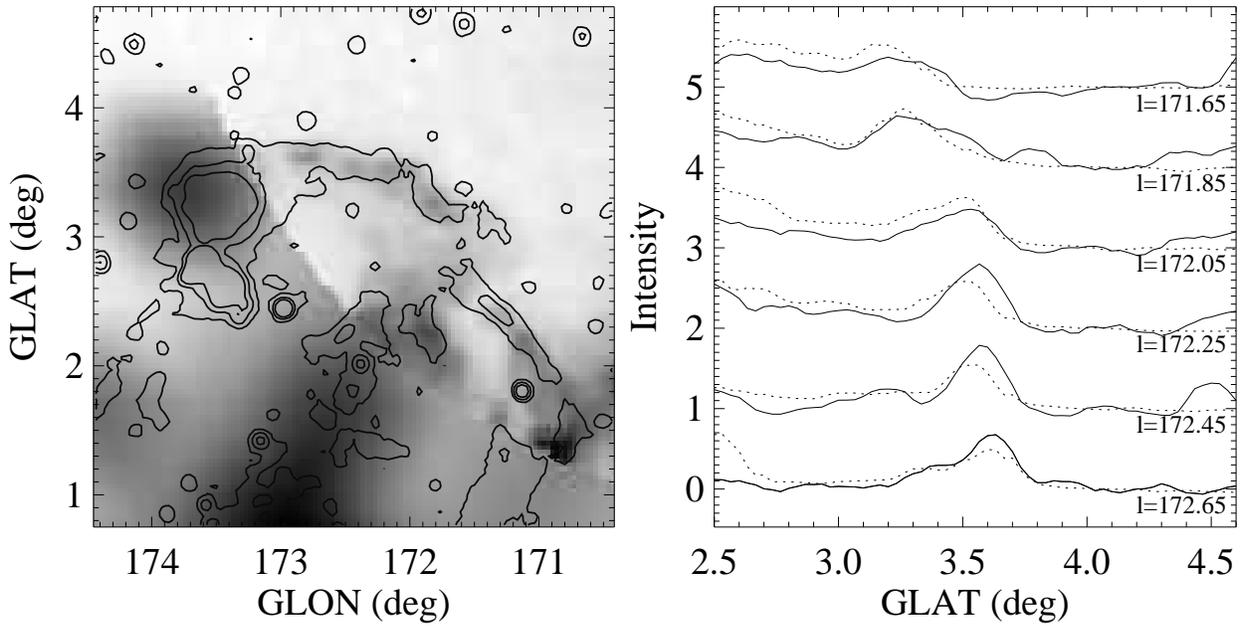}
\caption{(Left) An enlarged view of the filament A in
the VTSS \halpha image with the 11-cm radio continuum contours superposed.
(Right) One-dimensional intensity distributions of H$\alpha$ (dotted lines)
and 11-cm continuum (solid lines) emission along Galactic latitudes at
several Galactic longitudes. H$\alpha$ and continuum intensities are
normalized by 20 R and 0.1 K, respectively.
\label{fg10}}
\end{figure}

The radio continuum structure of G173+1.5 has counterparts in \halpha.
Fig.~\ref{fg9} (bottom left) is the H$\alpha$ full-sky map ($6'$ FWHM resolution)
composite of the Virginia Tech Spectral line Survey (VTSS), the Southern H$\alpha$ Sky Survey Atlas (SHASSA), and
the Wisconsin H-Alpha Mapper (WHAM) survey \citep{finkbeiner03}.
In the northern area covered by the VTSS \halpha image \citep{vtss}
delicate \halpha  filaments are
clearly visible, well correlated
with the radio continuum filaments. If we compare
in detail, however, there is a slight shift between
the radio and H$\alpha$ filaments;
the H$\alpha$ emission peaks at slightly lower latitudes, i.e.
slightly inside, the complex (Fig.~\ref{fg10}).
If the radio feature is due to thermal emission, the ratio of the
two intensities are roughly constant as both depend on
the emission measure. Using the formulae given in
 \citet{spitzer78}, the H$\alpha$ intensity from
ionized gas at 10,000 K is given by
$I(H\alpha)\approx$ 0.36$(n_p/n_e)EM $ Rayleigh (R)
where $EM =\int n_e^2 ds$ (cm$^{-6}$ pc), and
$n_p$ and $n_e$ are proton and electron densities, respectively.
The corresponding radio brightness temperature at 11 cm, using the formulae
from \citet{draine11}, is
$T_b\approx 0.40\times 10^{-3} (n_p/n_e)EM$ K.
Therefore, we have $T_b/I(H\alpha)\approx4\times 10^{-3}$ K R$^{-1}$
adopting an extinction of $A_V=1.4$ mag. At $T=5,000$ K,
this ratio would be slightly lower.
Since the H$\alpha$ intensity is normalized by 20 R and
the 11-cm brightness is normalized by 0.1 K,
the thermal 11-cm emission would have a
comparable ($\sim 0.8$) normalized intensity in Fig.~\ref{fg10}.
In fact, Fig.~\ref{fg10} shows that at the H$\alpha$ peak positions this is
indeed the case. On the other hand, at the radio continuum
maxima, the 11-cm brightness is considerably greater than
the H$\alpha$ intensity, which implies a non-thermal
origin. Therefore, the 11-cm continuum emission
appears to be composed of both thermal and non-thermal
components, which is consistent with our conclusion
from the previous section.

Figure~\ref{fg9} (bottom right) show the CO J=1-0 intensity map
integrated over \vlsr$=-25$ to $-15$~\kms
\citep {dame01}.
There is one giant molecular cloud at
$(\ell,b)\sim (173.^\circ5,2.^\circ5)$.
Four \schii regions, S231, S232, S233, and 235,
are clustered around this molecular cloud.
CO observations of this cloud show close
positional correlations between the optical \schii regions and the molecular cloud
\citep{heyer96}.
The giant molecular cloud appears as interconnected filaments with bright CO
emission located at the projected optical edges of S235 and S232. S233 lies
within a void in the CO emission, implying complete ionization and
photodissociation of the molecular material.
It has been known that active star formation is ongoing around
these \schii regions, and that jets and outflows from young stellar objects have been observed
in molecular lines
\citep[e.g., ][]{shepherd02, beuther02}.

\section{Discussion}

\subsection{The Origin of FVW172.8+1.5}

Our results indicate that the fast-moving \schi gas in
FVW172.8+1.5 is confined within the HII complex G173+1.5 and
has good spatial correlation with the radio continuum structure.
This strongly suggests its origin lies within the complex.
Morphological association between
the \schi emission and the continuum/H$\alpha$ filaments further suggests
that they probably trace the same object, i.e., an expanding shell.
The X-rays detected within the complex may also be associated
with this shell.
In this section, we first derive the parameters
that an expanding shell would possess, and then explore its possible origin.
We take the systemic velocity and distance of both the HII complex
and FVW172.8+1.5 to be
$v_{\rm sys} \approx -20$~\kms and $1.8$ kpc.

\subsubsection{The parameters of FVW172.8+1.5 as an expanding shell}

The total extent of the putative shell based on the radio continuum
filaments and our \schi observations is $4^\circ.4\times 3.^\circ.4$,
which converts to $138\ {\rm pc} \times 107$ pc at 1.8 kpc.
If we adopt the velocity centers of the highest-velocity clumps
given in \S~4 (+35~\kms) as an endcap velocity of the shell,
its expansion velocity is 55~\kms.
To derive the total \schi mass and kinetic energy of the FVW172.8+1.5,
we need to estimate the {\em total} mass of the expanding shell.
This can be done by extrapolating the mass distribution we derive for
high velocities. Fig. 11 shows the distribution of \schi
mass in each 2.2~\kms velocity interval
between \vlsr = +14 and +46~\kms.
We assume the receding and approaching hemispheres of the expanding
shell to be symmetric and centered at \vlsr $= -20$~\kms.
Note that the approaching portion of the shell is not visible 
because of confusion by the Galactic background/foreground \schi emission.
While this may not be true, the resulting kinetic energy would
be roughly correct should the energy injection be
spherically symmetric.
A Gaussian
shape is often used for the extrapolation
when fitting the mass distribution of observed \schi shells
\citep[e.g.][]{giovanelli79, koo90, kks06}.
If we adopt this Gaussian extrapolation,
(the solid line in Fig. 11), we
obtain a value of $1.3\times 10^4 d_{1.8 {\rm kpc}}^2$\ms.
However, this may overestimate the total mass because
for an ideal expanding thin shell of uniform density,
the mass per unit radial velocity interval is constant.
The mass profile at the highest velocities could be decreasing
due to the velocity dispersion, or the velocity structure, within
the shell.
However, the mass profile at the central velocities would remain constant
unless the turbulent velocity in the shell were large enough to be
comparable to its expansion velocity.

Here, we develop a simple, but physical, shell model
that yields a flat mass profile for the central velocities.
We consider a spherical shell with a radius $R=$2\degr, a thickness 
$\Delta R=10$\arcmin, an expansion velocity $v_{\rm exp} = 55$~\kms,
and a dispersion velocity $v_\sigma$ = 5.5~\kms.
These parameters are derived from the observed \schi shell.
We assume that the gas density within the shell is constant.
For the radial profile of expansion velocity within the shell,
we consider two different cases.
The first is that the expansion velocity is constant with radius. The second
is that the expansion velocity varies linearly
with radius. Linearly decreasing or increasing velocity structures
within the shell are derived by theoretical studies for different ages of
radiative shells \citep{slavin92, blondin98}.
Here,
we choose a velocity structure for which the expansion velocity at the
outer radius drops to 50\% of that at the inner radius. 

To fit the derived mass distribution,
we adopt a three dimensional cube composed of 451$^3$ pixels. The $z$
axis is along the line of sight (LOS). The number density of \schi atoms
for a pixel centered at $r=(x^2+y^2+z^2)^{1/2}$ is $n(r)$,
which is assumed to be constant in our model.
This pixel has an LOS velocity
$v_{los}(x, y, z) =v_{\rm exp}(r)z/r$ where $v_{\rm exp}(r)$
is the radial expansion
velocity at $r$. The \schi
21-cm emission line from this pixel has a Gaussian distribution with the LOS
velocity dispersion of $v_\sigma$ (=5.5~\kms).
Then the column density, $N(x, y, v)$,
at $(x,y)$ over the LSR velocity interval from $v$ to $v+\Delta v$ and the
\schi mass, $M(v)$, between $v$ and $v+\Delta v$, are given as follows:
\begin{displaymath}
N(x, y, v)  = {n\over \sqrt{2 \pi v_\sigma^2}}\int
\int_v^{v+\Delta v} \exp\left[{-{(v^\prime -
v_{los}(x, y, z))^2}\over {2 v_\sigma^2}}\right] dv^\prime dz,
\end{displaymath}
\begin{displaymath}
M(v) = m_{\rm H} \int \int N(x, y, v) dxdy.
\end{displaymath}

\begin{figure}
\epsscale{0.8}
\plotone{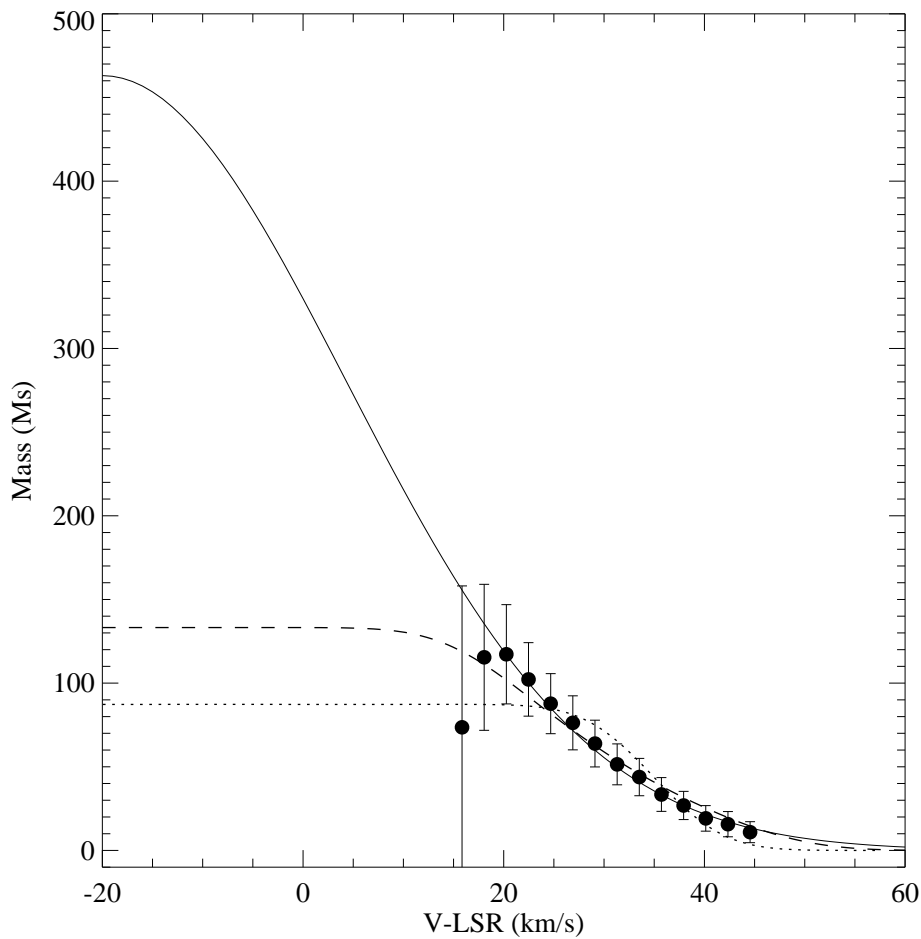}
\caption{
The mass distribution for the putative \schi shell assuming a distance of 1.8 kpc.
The filled circles are the derived \schi masses for each 2.2 \kms velocity interval.
The error bars show the standard deviation for the background area.
The solid line shows a Gaussian fit to the data, while the dotted line
shows the best fit for the mass distribution of a shell
with a constant \vexp=55~\kms and $v_{\sigma}= 5.5$ km s$^{-1}$,
and the dashed line shows
the best fit for
the mass distribution of a shell with a velocity dispersion
$v_{\sigma}=5.5$ km s$^{-1}$, but whose expansion
velocity at the outer radius of the shell drops linearly to 50\% of the
expansion velocity at the inner radius.
\label{fg11}}
\end{figure}

The dotted line in
Fig.~\ref{fg11} is the best fit for the mass distribution of a shell with
a constant expansion velocity profile.
Here, the only fitting parameter is the density of the shell.
The velocity range, where the mass of the observed \schi shell
smoothly decreases,
is wider than the velocity range due to velocity dispersion, so the
observed \schi mass distribution cannot be explained by velocity
dispersion alone.
The dashed line shows the best fit for the mass distribution of a shell
with an expansion velocity profile linearly decreasing
towards larger radii.
This dashed line fits the observed mass better
than the dotted line.
Variation of
the expansion velocity with radius dilutes the rectangular mass distribution
at high velocities, but still yields
a constant mass distribution at lower velocities,
from \vlsr $\sim-50$ to $+10$~\kms.
Note that the basic shape of the mass profile is determined by the difference
between the minimum and maximum expansion velocities.
The detailed velocity and density
structures only affect the shape of the decreasing portion of the mass
profile. We will study the mass profiles resulting from various
velocity and density structures in a forthcoming paper.
The estimated
total \schi mass of the shell adopting a linearly decreasing velocity profile
is 5900$d_{1.8 {\rm kpc}}^2$\ms.
We will use this value as the characteristic mass of the shell. Note that
the total mass derived from the Gaussian fit is about twice this.
The kinetic energy of the shell, taking into account the He abundance of 10\%
by number, is $0.25\times10^{51} d_{1.8 {\rm kpc}}^2$ ergs.
The derived physical
parameters of the model \schi/continuum shell are presented in Table~\ref{tbhi}.

\subsubsection{The Origin of FVW172.8+1.5 and the X-rays}

We can think of two possibilites for the origin of FVW172.8+1.5;
stellar winds from OB stars or SN explosions.
HII regions can also produce expanding \schi shells, but their
expansion velocities and energies are much smaller than those of
FVW 172.8+1.5.
Furthermore, the \schii regions
are mostly located at the boundary of the structure, so that they
do not appear to be major sources of the $3^\circ$--$4^\circ$-sized shell.
We first consider whether the stellar winds from OB stars can produce
FVW172.8+1.5.
There are more than a dozen O stars in this area \citep{garmany82,maiz04}
with eight of them projected against the shell (Figs. 5 and 9).
We summarize the parameters of these stars in Table 6. Their spectral types
range from O9.5 to O7,
and four are associated with the Sharpless HII regions.
The other four
could be possible candidates for the origin of the {\sc Hi}/continuum shell.
If we use the relation of \citet{howarth89} and a
wind speed of $\sim 2,000$ \kms, then the typical wind luminosities of
O7--O9 main-sequence stars are (5--0.8)$\times 10^{35}$ erg s$^{-1}$.
For giant O stars, this would be larger by a factor of two.
However, the required wind luminosity to explain the
parameters of the shell is
$L_w = 77 E_K/(9 R/ v_{\rm exp}) = 6\times 10^{37}$ erg s$^{-1}$
\citep{weaver77}, which is two orders of magnitude greater than
the available wind luminosities.
Even if there are groups of stars in this area (see next section), their
contribution to the wind luminosity
will not be significant because they are no brighter than O9.

We thus consider that FVW172.8+1.5 is likely to be the result of
an SN explosion.
Certainly, the derived parameters are fully consistent with this interpretation.
The required SN explosion energy is
$E_{\rm SN} = 6.8 \times 10^{43} n_0^{1.16}
{R_s}^{3.16} {v_{\rm exp}}^{1.35} \chi^{0.161}$~ergs,
where $n_0$ is the ambient H density in cm$^{-3}$, $R_s$ is the radius of the shell
in pc, $v_{\rm exp}$ is the expansion velocity in km s$^{-1}$,
and $\chi$ is the metalicity in units of the solar metalicity \citep{cioffi}.
For the above parameters, $E_{\rm SN}=1.3 \times 10^{51}$ ergs,
where the solar metalicity has been adopted.
This value of $E_{SN}$ is close to the canonical outburst
energy of a single SN explosion of $1\times 10^{51}$ ergs.
The age of the shell, assuming this to be in the radiative stage of SNR
evolution, is $t$ = 0.3 $R_{\rm s} / v_{\rm exp}$ = 0.33 Myr.
The SNR interpretation is also consistent with the presence of
X-rays in this region. The thermal energy of the X-ray emitting gas
$3\times 10^{50}$~ergs, which is 25\% of the initial explosion energy
and close to that expected for a 0.33 Myr-old radiative SNR \citep[e.g.,][]{cioffi}.
The apparent segregation between the fast-moving \schi gas and
the X-ray emitting gas could be due to
a non-uniform and/or inhomogeneous density distribution in the ambient medium.
It is also possible that more than one SN exploded within the
past few $10^{5}$ yr.
We have checked for known pulsars in this area,
but none are known within the boundary of the putative shell.
The nearest pulsar, PSR
J0540+32 at ($\ell,b$) = ($176.\!\!^{\circ}$80, +0.$\!\!^{\circ}$72),
is $\sim 4^\circ$ distant from the
center of the {\sc Hi}/continuum feature, and lies at a distance of $\sim$6.2
kpc. PSR J0540+32 is thus unlikely to be associated with the
progenitor of FVW172.8+1.5.

\subsection{Speculations on the Star-Formation History of G173+1.5}

The HII complex G173+1.5 is composed of five
Sharpless \schii regions S231--235. Of these,
all but S234 are clustered around
a giant molecular cloud (Fig. 9).
There are at least fourteen
embedded star-forming clusters that are 3--5 Myr old
around these four HII regions
\citep[][and references therein]{kirsanova08, dewangan11,camargo11}.
S234, located $3^\circ$ below the other HII regions,
also has an associated cluster with young stellar objects
having an age of $\sim 3$ Myr \citep{jose08}.
It has been proposed that the formation of some of these star-forming clusters
could have been triggered by the first-generation stars in these HII regions.

The presence of an old SNR filling the HII complex G173+1.5
has an interesting implication.
The five Sharpless HII regions are located within the
boundary of the {\sc Hi}/continuum structure,
suggesting that their
formation could have been triggered by SN explosions,
stellar winds, or expanding HII regions from a previous generation of stars
\citep[See][for a review of triggered star formation]{elmegreen98}.
The age of the SNR for which we have found evidence
is, however, only 0.33 Myr, so that
it is not possible that the current expanding shell
triggered the formation of these HII regions.
Instead it could be the cluster to which
the SN progenitor belonged that triggered the formation of
these HII regions.
There are two stellar associations in this area; Aur OB1 at a distance
of 1.3 kpc and Aur OB2 at 3.2 kpc \citep{humphreys78}.
We suspect that Aur OB1, which is
scattered over a wide area,
($\ell=168.^\circ1$--$178.^\circ1$, $b=-7.^\circ4$--$4.^\circ2$),
could be
composed of several associations at slightly different distances.
The distance moduli of Aur OB1 stars indeed range from 9.69 to 11.67
or 0.9 to 2.2 kpc in distance \citep{humphreys78}.
HD 36483 (O9IV) and HD 35921 (O9.5III)
that are relatively close to the central area of the {\sc Hi}/continuum structure
also belong to Aur OB1.
If one, or both, of these also belonged to the
appropriate cluster, then since the main-sequence lifetime of a $20$~$M_\odot$ O9 star
is $\sim 8$ Myr \citep{schaller92}, the star formation in this area
began about $\simgt 8$~Myr ago. In this scenario,
the first generation stars triggered the
formation of a second generation that are currently
exciting the Sharpless HII
regions (S231-S235), and we are now observing the formation of third generation
stars around these HII regions.
The HII complex G173+1.5 could be another good example
of sequential star formation over several stellar associations.

\section{Summary}

We have made
high resolution \schi 21-cm line observations of a $7~\!\degr\!\!.5 \times
4~\!\degr\!\!.5$ area centered at $(\ell,b) = (173\degr,-1.5\degr$)
using the ALFA 7-beam receiver mounted on the Arecibo 305~m telescope.
The data cube provides a detailed view of the fast-moving \schi gas
at velocities forbidden by Galactic rotation in this area.
FVW 172.8+1.5 is a feature that appears as a faint wing in low-resolution
surveys. By comparison with radio continuum, X-ray, and H$\alpha$ maps, we find
evidence that
FVW 172.8+1.5 is likely to be an old SNR within a large HII complex
situated in this area.
Our results are summarized as follows:

\noindent
(1) The \schi emission associated with FVW172.8+1.5 is visible
from \vlsr=+19.5 to +50.4~\kms.
At the highest velocities, the {\sc Hi} appears to be separated
into two concentrations. At lower velocities
a thin, filamentary feature appears surrounding these two concentrations.
The increase of the surface area and the amount of
\schi gas with decreasing velocity suggests that it is
an expanding shell. We attribute the complex
morphology to the non-uniformity and
inhomogeneity of the ambient medium.

\noindent
(2) The high-velocity \schi emission features
are confined inside the radio continuum filaments associated with
the large HII complex G173+1.5
and show a very good spatial correlation with them. This
strongly suggests that it has an origin within the complex.
Adopting a
distance of 1.8 kpc, and the systemic velocity of $-20$ \kms of the HII 
complex to be appropriate for
the \schi shell, we derive physical parameters for the shell.
These yield a kinetic energy of $2.5\times 10^{50}$~ergs.
This large kinetic energy and the fast-expansion velocity implies a SN origin.

\noindent
(3) We investigated the nature of the filaments
in HII complex G173+1.5 using multi-wavelength
data. Our analysis of the radio continuum spectrum
suggests that the radio filaments have
both thermal and non-thermal components. A detailed comparison with
the associated H$\alpha$ filament supports this conclusion.
We also find X-ray emitting
hot gas inside the complex with an estimated thermal energy of $3\times 10^{50}$ ergs.
This associated hot gas of comparable thermal energy to  the kinetic energy of
expansion, and
the probable presence of non-thermal radio filaments, supports the conclusion that
the HV \schi gas is of SN origin.

\noindent
(4) From our analysis, we conclude that FVW 172.8+1.5 is likely to be an
old ($\sim 0.33$ Myr) SNR produced inside the HII complex G173+1.5.
We propose that the stellar association to which the progenitor belonged
could have triggered the formation of the OB stars currently
exciting the HII regions. The HII complex G173+1.5 appears to be
a potential example of sequential
star formation over several stellar associations.

\acknowledgments
We wish to thank Josh Goldston, Carl Heiles and others who provided
enormous help with the observations and data reduction. 
We are grateful to Min Jin Kim for his help with the continuum data analysis.
We thank Tom Dame for kindly providing the CO data cube.
Ed Churchwell is thanked for many useful discussions.
We also wish to thank the referee, Tom Landecker, for his helpful comments.
We would like to thank the staff of the Arecibo Observatory
for enormous help with the observations and data reduction.
This work has been supported by the Korean Research
Foundation under grant KRF-2008-313-C00372 to B.-C. K.
The Arecibo Observatory is part of the National Astronomy and
Ionosphere Center, which is operated by SRI International under a
cooperative agreement with the National Science Foundation
(AST-1100968), and in alliance with Ana G. Mndez-Universidad
Metropolitana, and the Universities Space Research Association.  The
Virginia Tech Spectral-Line Survey (VTSS) is supported by the National
Science Foundation.

\begin{deluxetable}{ccrrrr}
\tabletypesize{\small}
\tablecaption{Sharpless \schii regions in the observed field\label{tbl-1}}
\tablewidth{0pt}
\tablehead{
\colhead{} & \colhead{RA, Dec (J2000)} & \colhead{Diameter\tablenotemark{a}} &
\colhead{\vlsr\tablenotemark{b}} & \colhead{\vlsr\tablenotemark{c}} &
\colhead{Distance\tablenotemark{d}} \\
\colhead{Name} & \colhead{( h m, $\degr$ $^\prime$ )} & \colhead{(\,\arcmin\,)} &
\colhead{(km s$^{-1}$)} & \colhead{(km s$^{-1}$)} & \colhead{(kpc)} \\
}
\startdata
S229 & 5 16.3, +34 27 &   65&      $+4.4$&    $+6.7$&    0.51    \\
S230 & 5 22.5, +34 08 & 300 &    $+0.0$ &     ... & ... \\
S231 & 5 39.3, +35 56 &   12&      $-17.5$&   $-18.1$&   2.3      \\
S232 & 5 42.5, +36 12 &   40&      $-13.7$&   $-23.0$&   1.0       \\
S233 & 5 38.7, +35 48 &   2 &      $-14.5$&   $-18.4$&   ...  \\
S234 & 5 28.1, +34 26 &   12&      $-14.3$&   $-13.4$&   2.3      \\
S235 & 5 41.0, +35 51 &   10&      $-25.7$&   $-18.8$&   1.6      \\
S236 & 5 22.6, +33 22 &   55&      $-3.8$&    $-7.2$&    3.2      \\
S237 & 5 31.4, +34 17 &    7 &      +1.4 &    $-4.3$&    1.8  \\
\enddata
\tablenotetext{a}{Optical size \citep{sharpless59}}
\tablenotetext{b}{\halpha emission line radial velocity \citep{fich90}}
\tablenotetext{c}{CO radial velocity \citep{blitz82}.
For S229, the association is uncertain. }
\tablenotetext{d}{Distance determined by spectrophotometry
of the central exciting star. See the references in \citet{blitz82}.}
\end{deluxetable}

%

\begin{deluxetable}{cccrl}
\tabletypesize{\small}
\tablecaption{Parameters of the continuum surveys used for spectral analysis \label{tbcont}}
\tablewidth{0pt}
\tablehead{
\colhead{} & \colhead{Frequency} & \colhead{} & \colhead{RMS} &
\colhead{} \\
\colhead{Name} & \colhead{(MHz)} & \colhead{Resolution} & \colhead{(mK)} &
\colhead{References}\\}
\startdata
WENSS &325     &54~\arcsec $\times 54$~\arcsec csc $\delta$  & 2300
& \citet{reng97}\\
CGPS &408  &2~\arcmin$\!\!.8\times 2\arcmin\!.8$ csc $\delta$   & 950
&\citet{taylor03}\\
CGPS &1420 &49~\arcsec $\times 49$~\arcsec csc $\delta$   & 68
&\citet{taylor03}\\
Effelsberg 11-cm  &2695    &$4\arcmin\!.3$    &20 & \citet{furst}\\
\enddata
\end{deluxetable}

\begin{deluxetable}{llcccc}
\tabletypesize{\small}
\tablecaption{Integrated radio continuum flux densities of Filament A and S232 \label{tbtotflux}}
\tablewidth{0pt}
\tablehead{
\colhead{} & \colhead{325 MHz} & \colhead{408 MHz} &
\colhead{1420 MHz} & \colhead{2695 MHz} & \colhead{Spectral index}\\
\colhead{} & \colhead{(Jy)} & \colhead{(Jy)} &
\colhead{(Jy)} & \colhead{(Jy)} & \colhead{($S \propto \nu^{-\alpha}$)}}
\startdata
Filament A& $1.5\pm0.17$ & $0.84\pm0.14$ & $0.87\pm0.03$ &$0.73\pm0.05$
&$+0.23\pm 0.06$\\
S232& $1.8\pm0.07$ & $1.9\pm0.10$ & $1.9\pm0.10$ &$2.2\pm0.05$
&$-0.09\pm 0.02$\\
\enddata

\tablecomments{The areas used for the flux density measurements of the filament A
and S232 are indicated in Fig.~\ref{fg7}.}
\end{deluxetable}

\begin{deluxetable}{lcccc}
\tabletypesize{\small}
\tablecaption{Spectral indices from T-T plots \label{tbtt}}
\tablewidth{0pt}
\tablehead{
\colhead{} & \colhead{325-1420 MHz} & \colhead{325-2695 MHz} &
\colhead{408-1420 MHz} & \colhead{408-2695 MHz}}
\startdata
Filament A & $+0.50\pm0.11$ & $+0.32\pm0.06$ & $+0.07\pm0.10$ & $-0.04\pm0.08$\\
S232 & $+0.09\pm0.07$ & $-0.00\pm0.04$& $+0.06\pm0.04$& $-0.04\pm0.02$\\
\enddata
\end{deluxetable}

\begin{deluxetable}{ll}
\tabletypesize{\small}
\tablecaption{Parameters for an expanding shell\label{tbhi}}
\tablewidth{0pt}
\tablehead{
\colhead{Parameter} & \colhead{Estimated value}\\}
\startdata
Center $(\alpha, \delta)$ & $05^{\rm h} 33^{\rm m} 29^{\rm s}, +35\degr 50\arcmin 52\arcsec$ \\
~~~~~~~$(\ell, b)$      & $172\degr.78, +1\degr.51$ \\
Mean radius             & $69 \times 54d_{1.8 {\rm kpc}}$  pc\\
Expansion velocity      & $55 \pm5$~\kms \\
Total \schi mass        & $5900 \pm 770$ $d_{1.8 {\rm kpc}}^2$ \ms \\
Kinetic energy          & $0.25 \pm 0.05 \times 10^{51}$ $d_{1.8 {\rm kpc}}^2$ ergs\\
Initial ISM density     & $0.25 \pm 0.03$ $d_{1.8 {\rm kpc}}^{-1}$ cm$^{-3}$ \\
Age                     & $0.33 \pm 0.03$ Myr\\
\enddata
\tablecomments{The errors are $1\sigma$ errors. For the estimation of the 
initial ISM density and age, we use the geometrical average of the mean radius (61 pc).}
\end{deluxetable}

\begin{deluxetable}{lclcc}
\tabletypesize{\small}
\tablecaption{O-type stars within the HII complex G173+1.5\label{tbostar}}
\tablewidth{0pt}
\tablehead{
\colhead{} & \colhead{RA, Dec (J2000)} &
\colhead{Spectral} &
\colhead{Distance} & \colhead{Associated}\\
\colhead{Name} & \colhead{( h m, \,\degr\, \arcmin\,)} &
\colhead{Type} & \colhead{( kpc )} &
\colhead{ \schii Region}\\}
\startdata
HD 36483 & 5 33.8, +36 28 & O9IV& 1.3 & ...\\
HD 35921 & 5 29.8, +35 20 & O9.5III& 1.3 & ...\\
LS V+35 24 & 5 39.7, +35 53 & O9V & 2.1 & S231\\
BD +35 1201 & 5 41.0, +35 50  &O9.5V & 1.6 &S235\\
HD 37737 & 5 42.4, +36 09 & O9.5III &1.3 & S232\\
BD +34 1058 &5 28.4, +34 40 &O8  &3.2 & ...\\
HD 35619 & 5 27.5, +34 47 & O7V & 3.2 & ...\\
BD +34 1054 & 5 28.1, +34 25 & O9.5V &  2.0 & S234\\
\enddata
\tablecomments{The O stars \citep{garmany82, maiz04}
located within a radius of $2^\circ5$ from the
center of the {\sc Hi}/continuum structure
are listed in order of angular distance from the center.
See also Fig.~\ref{fg5}.
The distance of a star is computed from
its distance modulus, assuming an interstellar
absorption of A(V)=3E(B-V)
\citep{garmany82}. }
\end{deluxetable}

\end{document}